\documentclass[conference]{IEEEtran}
\IEEEoverridecommandlockouts
\usepackage{cite}
\usepackage{amsmath,amssymb,amsfonts}
\usepackage{algorithmic}
\usepackage{graphicx}
\usepackage{textcomp}
\usepackage{xcolor}
\usepackage{mathtools}
\usepackage{amssymb}
\usepackage{braket}
\usepackage{nicefrac}
\usepackage{xfrac}
\usepackage{caption}
\usepackage{subcaption}
\usepackage{multicol}
\usepackage{float}
\def\BibTeX{{\rm B\kern-.05em{\sc i\kern-.025em b}\kern-.08em
    T\kern-.1667em\lower.7ex\hbox{E}\kern-.125emX}}
\begin{document}

\title{Quantum-Assisted Graph Clustering and Quadratic Unconstrained D-ary Optimisation\\
}

\author{\IEEEauthorblockN{Sayantan Pramanik and M Girish Chandra}
\IEEEauthorblockA{\textit{TCS Research and Innovation} \\
\textit{India}\\
sayantan.pramanik, m.gchandra@tcs.com}

}

\maketitle

\begin{abstract}
Of late, we are witnessing spectacular developments in Quantum Information Processing with the availability of Noisy Intermediate-Scale Quantum devices of different architectures and various software development kits to work on quantum algorithms. Different problems, which are hard to solve by classical computation, but can be sped up (significantly in some cases) are also being populated. Leveraging these aspects, this paper examines unsupervised graph clustering by quantum algorithms or, more precisely, quantum-assisted algorithms. By carefully examining the two cluster Max-Cut problem within the framework of quantum Ising model, an extension has been worked out for max 3-cut with the identification of an appropriate Hamiltonian. Representative results, after carrying out extensive numerical evaluations, have been provided including a suggestion for possible futuristic implementation with qutrit devices. Further, extrapolation to more than 3 classes, which can be handled by qudits, of both annealer and gate-circuit varieties, has also been touched upon with some preliminary observations; quantum-assisted solving of Quadratic Unconstrained D-ary Optimisation is arrived at within this context. As an additional novelty, a qudit circuit to solve max-d cut through Quantum Approximate Optimization algorithm is systematically constructed.
\end{abstract}


\section{Introduction}\label{introduction}
We are progressing through an exciting period in Quantum Technologies and with the small-scale commercial quantum computers becoming increasingly available \cite{2018arXiv181203050T}, Quantum Information Processing is witnessing spectacular developments.   Before quantum processors become scalable devices capable of error correction and universality \cite{kim2019leveraging}, the current and near-term devices, referred to as the Noisy Intermediate-Scale Quantum (NISQ) \cite{preskill2018quantum} devices are getting explored for solving certain hard problems to achieve significant speedups over the best known classical algorithms \cite{1997quant.ph..1001B}.  Promising results are already reported for solutions in the areas of optimisation, chemistry, machine learning. Apart from only speed up considerations, exploiting quantum mechanical properties of superposition, entanglement and interference for solving problems differently with possible performance improvements are getting explored, among others.  Needless to say, hybrid quantum algorithms which use both classical and quantum resources to solve potentially difficult problems \cite{zhu2019training} are worked out and put into action. 

It has been brought out that unsupervised machine learning and the associated optimisation strategies can be elegantly handled by quantum or hybrid quantum algorithms.  In this paper, we consider clustering, an important unsupervised task.  Clustering consists of assigning labels to elements of a dataset based only on how similar they are to each other - like objects will have the same label, unlike objects will have different labels \cite{otterbach2017unsupervised}. In order to represent dissimilarity (or similarity), we need to define a distance measure between  two  data samples.  The distance between every possible  pair  of  data  samples can be captured in a matrix. This matrix can be interpreted as an adjacency matrix of a graph, where each vertex or node represents an element of data set and the the weight of edge between vertices is the corresponding distance \cite{otterbach2017unsupervised}. In clustering, the main assumption is that distant points belong to different clusters; hence maximizing the overall sum of all weights (distances) between nodes with different labels represents a natural clustering algorithm for two-cluster case. The mathematical formulation of this is a well known Maximum-Cut (Max-cut) problem and it can be easily translated to an optimisation objective \cite{otterbach2017unsupervised}. The Max-cut problem  is  an  example  of  the  class of  NP-complete  problems,  which  are  notoriously hard to solve.  The Max-Cut and many other combinatorial problems, like, machine scheduling, computer-aided design, traffic message management \cite{otterbach2017unsupervised} fall under the unifying model of Quadratic Unconstrained Binary Optimisation (QUBO) \cite{otterbach2017unsupervised, lewis2017quadratic}.  One approach to solving Max-cut is to construct a physical system, typically a set of interacting spin  particles (two-state particles) whose lowest energy state encodes the solution to the problem, so that solving the problem is equivalent to finding the ground state of the system.

Two main approaches have been identified to find the ground state of interacting spin systems (quantum optimisation) in NISQs \cite{kim2019leveraging,otterbach2017unsupervised}: Quantum Annealing (QA) and Quantum Approximate Optimisation Algorithms (QAOA) \cite{farhi2014quantum}. QA is a form of analog computation that has been developed theoretically in the early nineties but realized experimentally in a programmable device only in 2011 by D-Wave Systems.  QAOA, invented in 2014 and recently generalised for constrained combinatorial optimisation, requires digital gate-model quantum computing; it can be seen in some parameter range as a “digitised” version of QA \cite{kim2019leveraging}.

In this paper, our starting point is a graph and we examine clustering on this abstraction (the graph itself can be constructed from the data points as cursorily mentioned in the beginning). To start with, we briefly touch upon the 2-cluster max-cut problem in terms of the usual Ising model of interacting spins, but report some additional results/observations related to graph components (independent subgraphs). Then,  we propose a simple way to extend the strategy to address 3-cluster problem on graphs.  The requisite 3-state particles interaction and the associated Hamiltonian are brought out. Apart from providing typical results, remarks on how to go about implementations are also made, including on the hypothetical qutrit computing device.  Extrapolation to more than three cluster case involving qudits is also suggested, culminating in the Quadratic Unconstrianed $D$-ary Optimisation (QUDO).

The paper is organized as follows: In Section \ref{maxcut}, Ising model and the max-cut clustering are presented.  Clustering into $3$ classes is covered in detail in Section \ref{max3cut}. Graph clustering into $d$ classes is brought out in Section \ref{maxkcut}.  In both Section \ref{max3cut} and Section \ref{maxkcut} relevant results are interspersed. Remarks related to implementation are provided in Section  \ref{implementation} including systematic steps for constructing the qudit-based circuit for $d$-ary QAOA to solve max $d$-cut. Conclusions are provided in Section \ref{conclusions}.     

\section{Two-Group Clustering Max-Cut Problem}\label{maxcut}
As mentioned in the previous section, one way to solve the two-cluster graph maxcut problem is to have a model of two-state interacting particles and solve for the lowest energy state.  This interaction model (for spins) is the Ising model, originally developed to describe ferromagnetism, but subsequently extended to more problems \cite{1525}. 

\subsection{Ising Model}\label{IsingModel}

The Ising model can be formulated on any graph as follows: consider an undirected graph
$G=(V,E)$, where $V=\{v_{1},...,v_{N}\}$  is a set of $N$ sites, and $E$ is a set of edges representing the interactions between these sites. Every site $i$ has a corresponding spin variable $s_{i}$ \cite{kamenetsky2010ising}. These spins are binary-valued, taking values $+1$ for “up” or $-1$ for “down”. Two spins  $s_{i}$ and $s_{j}$  may interact with each other \cite{kamenetsky2010ising}. The energy of such an interaction depends on whether the values of the participating spins are the same or different: it is given by $J_{ij}s_{i}s_{j}$, where $J_{ij}$  is the strength of the interaction \cite{kamenetsky2010ising}. 

For each pair of interacting spins $s_{i}$  and $s_{j}$ (i.e., $J_{ij}\neq0$), there exists a corresponding edge $(i,j) \in E$. The state of the model, $\textbf{\emph{s}}$, is an assignment of all $N$ variables $s_{i}$, $1 \leq i \leq N$. The set of all possible configurations is $\xi = \{-1,1\}^N$ \cite{kamenetsky2010ising}. As well as pair-wise interactions, there can also be an external field that affects each site $i$ with energy $h_{i}s_{i}$. Thus, in the general case, the energy of a configuration $\textbf{\emph{s}} \in \xi$  is given by the so-called Edwards-Anderson Hamiltonian \cite{kamenetsky2010ising}:
\begin{equation}
	\label{eq:classicalIsing1}
	H(\textbf{\emph{s}})=\!\!\!\!\sum_{(i,j) \in E}\!\!\!\!J_{ij}s_{i}s_{j} + \sum_{i \in V}h_{i}s_{i}
\end{equation}

When $h_{i} = 0 \; \forall \; i \in V$, the system is said to have no external field (also called zero magnetic field condition), in which case the energy of the configuration $\textbf{\emph{s}}$ becomes:
\begin{equation}
	H(\textbf{\emph{s}})=\!\!\!\!\sum_{(i,j) \in E}\!\!\!\!J_{ij}s_{i}s_{j}
	\label{eq:classicalIsing}
\end{equation}

In this paper, we mostly consider the Hamiltonian with zero external field.  The system prefers lower energy states, i.e., those $\textbf{\emph{s}}$ that minimise $H(\textbf{\emph{s}})$.  An important task is to find configurations that minimise the energy of the system; such a configuration is known as ground state.

In order to obtain the Quantum Mechanical description of the Ising model of Equation \eqref{eq:classicalIsing1} and Equation \eqref{eq:classicalIsing}, one has to replace each $s_{i}$ by the Pauli-$Z$ matrix given by $\sigma_{z}^{i}$. As the state corresponding to the quantum mechanical interaction of one or more particles is given by the tensor products of the corresponding individual states, the $\sigma_{z}^{i} \sigma_{z}^{j}$ terms in $H(\boldsymbol{\sigma})$ denote the tensor product between $\sigma_{z}^{i}$ and $\sigma_{z}^{j}$, where $\boldsymbol{\sigma}$ is the cumulative spin configuration of the complete system. It must be noted that these terms need to be appropriately constructed through tensor products of $\sigma_{z}$ and $I$ matrices, as discussed in subsection \ref{maxcutIsing}, to capture the pair-wise interaction between the $i^{th}$ and $j^{th}$ spins. With these considerations in mind, the Equation \eqref{eq:classicalIsing1} and Equation \eqref{eq:classicalIsing} convert to \cite{atan}: 

\begin{equation}
	H(\boldsymbol{\sigma})=\sum_{ij}J_{ij} \sigma_{z}^{i} \sigma_{z}^{j} + \sum_{i}h_{i} \sigma_{z}^{i}
	\label{eq:quantumIsingHamiltonian}
\end{equation}

\begin{equation}
	H(\boldsymbol{\sigma})=\sum_{ij}J_{ij} \sigma_{z}^{i} \sigma_{z}^{j}
	\label{eq:finalQuantumIsingHamiltonian}
\end{equation}

\subsection{Max-cut Problem Using Ising Model}\label{maxcutIsing}
Little more formally, the max-cut problem starts with an undirected graph $G(V,E)$ with a set of vertices $V$ and a set of edges $E$ between the vertices. The weight $w_{ij}$ of an edge between vertices $i$ and $j$ is a positive real number, with $w_{ij} = 0$ if there is no edge between them. A cut is a set of edges that separates the vertices $V$ into two disjoint sets $V_{1}$ and $V_{2}$, such that $V_{1} \subseteq V$ and $V_{2}=V \setminus V_{1}$, and the cost of a cut is defined as the as the sum of all weights of edges connecting vertices in $V_{1}$ with vertices in $V_{2}$. One can connect the cut to the Ising model by identifying the vertices with $s_{i}$ and $w_{ij}$ with $J_{ij}$; $s_{i} = 1$ suggesting that node $i$ belongs to $V_{1}$ and $s_{i} = -1$ corresponds to $V_{2}$ (of course, $V_{1}$ and $V_{2}$ can be interchanged). The cost of the cut can, in fact, be written in terms of the objective function:
\begin{equation}
	C=\sum_{ij}\frac{w_{ij}}{2}(1-s_{i}s_{j})
	\label{eq:maxcutCost}
\end{equation}
The max-cut problem aims at partitioning the nodes in such a way that the cost of the resulting cut is maximised. On comparing Equation \eqref{eq:classicalIsing} and Equation \eqref{eq:maxcutCost}, it must be noted that the sign of the $s_{i}s_{j}$ term changes. Thus, the max-cut then boils down to finding the lowest energy state, instead of the maximum, of Ising model with $J_{ij} = w_{ij}$:

\begin{equation}
	\max C = \min H(\textbf{\emph{s}})
\end{equation}

\noindent by noting that scaling the cost function by a constant multiplicative factor does not change the optimal solutions. As remarked earlier, the Max-Cut problem is equivalent to QUBO formulation where the two-state variable is $0$ or $1$ binary valued instead of $1$ and $-1$; if the QUBO variables are denoted by $x_{i}$, the two formulations are related by $s_{i}=2x_{i}-1$ \cite{lewis2017quadratic,ushijima2017graph,gowda}.

The classical Ising max-cut can be extended to the quantum framework by replacing $s_{i}$ by $\sigma_{z}^{i}$, as discussed in Section \ref{IsingModel}. This gives us the final quantum Ising Hamiltonian, given by Equation \eqref{eq:IsingHamiltonian}, which needs to be minimised to get the optimum cut \cite{ushijima2017graph}. In carrying out the optimisation based on Equation \eqref{eq:IsingHamiltonian}, if two adjacent nodes fall into the same cluster, then $w_{ij}$ is added to the cost function, else it is subtracted.
\begin{equation}
	\label{eq:IsingHamiltonian}
	H(\boldsymbol{\sigma})=\sum_{ij}w_{ij} \sigma_{z}^{i} \sigma_{z}^{j}
\end{equation}

As discussed in the previous subsection, the $\sigma_{z}^{i} \sigma_{z}^{j}$ terms represent the interaction of the nodes through the use of tensor products, as shown in the following example. Let us consider a graph having $5$ vertices. If there is an edge between the nodes $v_{1}$ and $v_{3}$ having weight $w_{13}$, then the interaction between the respective vertices is given by:
\begin{equation}
	H_{13}=w_{13}(I \otimes \sigma_{z} \otimes I \otimes \sigma_{z} \otimes I )
\end{equation}
The Hamiltonian thus formed is always diagonal, as there are no $\sigma_{x}$ terms involved, owing to the tensor product between diagonal matrices. It is to be noted that Ising models can be of transverse field type, where, $\sigma_x$ terms corresponding to Pauli-X matrix also are present, in which case the corresponding Hamiltonian would be non diagonal.

\begin{figure}
	\centering
	\includegraphics[scale=0.59]{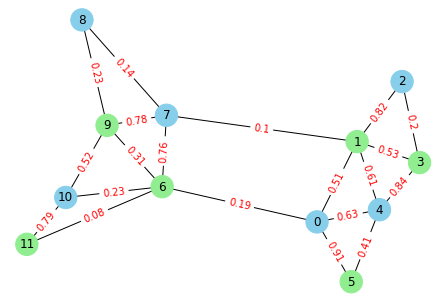}
	\caption{Result of max-cut clustering on weighted graphs. The colour of each node denotes the cluster it belongs to.}
	\label{fig:binary}
\end{figure}

The cost Hamiltonian $H(\boldsymbol{\sigma})$ of the graph $G$, being diagonal, has orthogonal eigenvectors that form a complete standard basis. The system settles to the state having the lowest energy and the eigenvector corresponding to it is given by a unit vector along a standard basis state. The least-energy eigenvector is a $2^{N} \times 1$ vector, whose ket representation gives us an $N$-length bit-string. The nodes of the graph, $G$, are labelled as $0$ or $1$, according to the digits in the bit-string, with the most significant bit representing the label of the first vertex. This gives us a binary-clustered graph. The result of partitioning a graph with the given algorithm has been shown in Figure \ref{fig:binary}. The minimum energy state corresponding to the partition is given by $\ket{010101100101}$.

\section{Solving Max 3-cut using Ising Model}\label{max3cut}
In the direction of arriving at the quantum-assisted solution for Max $3$-cut problem, few observations are put in place. Consider the nodes $2$ and $3$ of a graph with $6$ nodes, shown in Figure \ref{fig:2nodes} for binary clustering:

\begin{figure}[h!]
	\centering
	\includegraphics[scale=0.5]{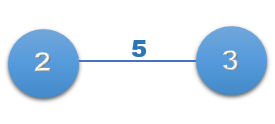}
	\caption{Two nodes connected by an edge}
	\label{fig:2nodes}
\end{figure}

\begin{figure}[h]
	\centering
	\includegraphics[scale=0.5]{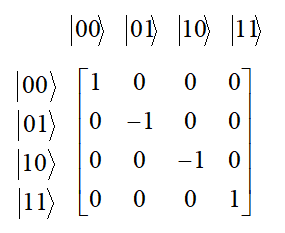}
	\caption{Node-node interactions and corresponding interaction energies}
	\label{fig:binaryHamiltonian}
\end{figure}

The cost for the connection, in the Ising model, is given by:
\begin{equation}
	H_{23}=5 \times (I \otimes I \otimes \begin{bmatrix} 1 & 0 \\ 0 & -1 \end{bmatrix} \otimes \begin{bmatrix} 1 & 0 \\ 0 & -1 \end{bmatrix} \otimes I \otimes I)
\end{equation}
which can be simplified as:
\begin{equation}
	H_{23}=5 \times (I \otimes I \otimes \begin{bmatrix} 1 & 0 & 0 & 0 \\ 0 & -1 & 0 & 0 \\ 0 & 0 & -1 & 0 \\ 0 & 0 & 0 & 1 \end{bmatrix} \otimes I \otimes I)
\end{equation}
The $4 \times 4$ matrix, in the equation above, incorporates all the possible cluster combinations of the nodes 2 and 3 (along the diagonal). If the two cluster are named $0$ and $1$, then their combinations and energies are shown in Figure \ref{fig:binaryHamiltonian}.

The energy is $1$ when both the nodes are classified into the same cluster, $\ket{00}$ or $\ket{11}$, and is $-1$ otherwise. The system will settle for the lower energy state, $-1$, thus providing the optimum clustering.

\begin{figure}[h]
	\centering
	\includegraphics[scale=0.65]{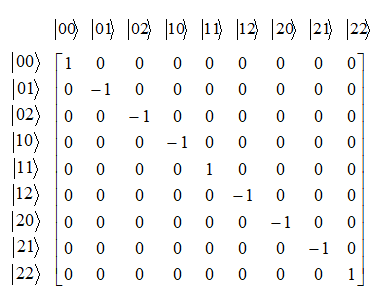}
	\caption{Node-node interaction energies for the ternary clustering case}
	\label{fig:ternaryHamiltonian}
\end{figure}

The idea can be extended to clustering the nodes of a graph into $3$ classes, with class labels $0$, $1$ and $2$. The interaction-matrix between adjacent nodes of the graph should ideally look like (coupling strength has been considered to be unity) the matrix in Figure \ref{fig:ternaryHamiltonian}.

In the binary case, the nodes were represented by $2 \times 2$ Pauli-$Z$ matrices, the tensor product of which gave the required Hamiltonian matrix. Thus, for the 3-class problem, we need $3 \times 3$ matrices for each node, whose tensor product with another such matrix can give the requisite $9 \times 9$ Hamiltonian. 

One can think of using a matrix with the cube roots of unity placed along the diagonal of the $3 \times 3$ matrix:
\begin{equation}
	\Omega_{3}=\begin{bmatrix} 1 & 0 & 0 \\ 0 & e^{\nicefrac{2 \pi i}{3}} & 0 \\ 0 & 0 & e^{\nicefrac{4 \pi i}{3}} \end{bmatrix}
\end{equation}

In an interaction between two nodes, the first node is represented by $\Omega_{3}$ and the second node is represented by the complex conjugate transpose of $\Omega_{3}$ , $\Omega_{3}^{\dagger}$ . The 3 in subscript signifies that the cube roots of unity are used to form the matrix. Thus, the energy of interaction of nodes 2 and 3 is:
\begin{equation}
	H_{23}=\dots \otimes \begin{bmatrix} 1 & 0 & 0 \\ 0 & e^{\nicefrac{2 \pi i}{3}} & 0 \\ 0 & 0 & e^{\nicefrac{4 \pi i}{3}} \end{bmatrix} \otimes \begin{bmatrix} 1 & 0 & 0 \\ 0 & e^{\nicefrac{-2 \pi i}{3}} & 0 \\ 0 & 0 & e^{\nicefrac{-4 \pi i}{3}} \end{bmatrix} \otimes \dots
\end{equation}
\begin{equation}
	\Rightarrow H_{23}=\dots \otimes \begin{bmatrix} 1 & 0 & \dots & 0 & 0 \\ 0 & e^{\nicefrac{-2 \pi i}{3}} & \dots & 0 & 0 \\ \vdots & \vdots & \ddots & \vdots & \vdots \\ 0 & 0 & \dots & e^{\nicefrac{2 \pi i}{3}} & 0 \\0 & 0 & \dots & 0 & 1 \\ \end{bmatrix} \otimes \dots
\end{equation}

But, the Hamiltonian, being an observable, must be Hermitian. This can be taken care of by modelling the interaction between two nodes as the tensor product between $\Omega_{3}$ and $\Omega_{3}^{\dagger}$ and taking only the real part of the elements of the resultant matrix. The final form of the interaction is given as:
\begin{equation}
	H_{23}=5 \times I \otimes I \otimes \frac{1}{2} (\Omega_{3}^{2} \otimes \Omega_{3}^{3 \dagger} + \Omega_{3}^{2 \dagger} \otimes \Omega_{3}^{3}) \otimes I \otimes I
\end{equation}

It is interesting to note that this is completely analogous to the $d=2$ case, where the Pauli-$Z$ matrix is Hermitian and thus, $\sigma_{z} \otimes \sigma_{z}^{\dagger} = \sigma_{z}^{\dagger} \otimes \sigma_{z}$.

The term $\bar{H}_{23}= \frac{1}{2} (\Omega_{3}^{2} \otimes \Omega_{3}^{3 \dagger} + \Omega_{3}^{2 \dagger} \otimes \Omega_{3}^{3})$ evaluates to:
\begin{equation}
	\bar{H}_{23}=\begin{bmatrix} 1 & 0 & \dots & 0 & 0 \\ 0 & -0.5 & \dots & 0 & 0 \\ \vdots & \vdots & \ddots & \vdots & \vdots \\ 0 & 0 & \dots & -0.5 & 0 \\0 & 0 & \dots & 0 & 1 \\ \end{bmatrix}
\end{equation}

\begin{figure}
	\centering
	\includegraphics[scale=0.65]{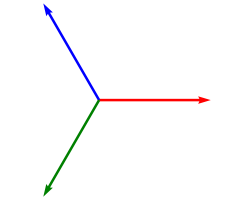}
	\caption{The cluster classes are modelled along the vectors: $0$ along red, $1$ along blue, $2$ along green.}
	\label{fig:ternaryVector}
\end{figure}

If adjacent nodes are placed in dissimilar clusters, the interaction energy is $-0.5$, and $1$ for similar clusters. This is in contrast to the desirable energy values of $-1$ and $1$ for dissimilar and similar clusters, respectively. But this does not have any effect on the clustering result since the energy for adjacent vertices in different clusters is still lower than that for similar clusters, and the former will be energetically favoured. To provide an easy visualisation, the cluster classes can be modelled along the cube roots of unity as shown in Figure \ref{fig:ternaryVector}, since the clustering energy can be considered in terms of angular separation, as discussed in the next section.

\begin{figure}
	\centering
	\includegraphics[scale=0.5]{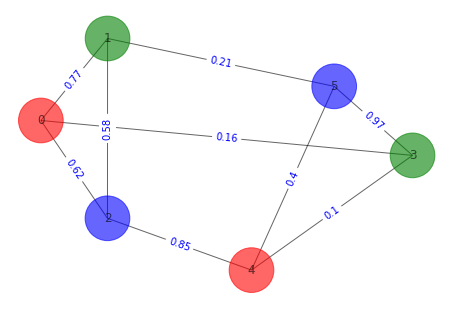}
	\caption{Results of max $3$-cut clustering where the three clusters are denoted by red, blue and green colours.}
	\label{fig:ternary}
\end{figure}

The typical result of max 3-cut partitioning on a graph can be seen in Figure \ref{fig:ternary}. The final Hamiltonian for max-cut is given by:

\begin{equation}
	\label{eq:max3cutHamiltonian}
	H=\sum_{ij}\frac{w_{ij}}{2}(\Omega_{3}^{i} \Omega_{3}^{j \dagger}+\Omega_{3}^{i \dagger} \Omega_{3}^{j})
\end{equation}

Extensive numerical evaluation studies with different graphs corroborates the applicability of the proposal.

\section{Max $d$-Cut}\label{maxkcut}

Having looked at max-cut and max $3$-cut, where the nodes of a graph are partitioned into two and three classes respectively, in this section, we propose a way of generalising the max-cut problem for $d$ classes. Such problems have traditionally been known as max $d$-cut \cite{Alan}. Before moving onto the problem, it is necessary to introduce some additional mathematical machinery to facilitate the solution.

In section \ref{max3cut}, the matrix $\Omega_{3}$ can be recognised as the clock matrix for three dimensions. This is not surprising as the clock and shift matrices, $U_{d}$ and $V_{d}$, have been used to generalise the Pauli-$Z$ and $X$ matrices \cite{prakash2018normal}, respectively, for $d$ dimensional \emph{qudits} \cite{Frydryszak_2017}. These matrices, which are zero-trace and symmetrical, but not Hermitian, can be utilised to solve $d$-ary optimisation problems where the variables can take $d$ number of discrete values, similar to a $d$-level system. In this section we extend the idea of Quadratic Unconstrained Binary Optimisation, $QU\!BO$, to such $d$-ary problems, and dub the technique as Quadratic Unconstrained $D$-ary Optimisation, $QU\!DO$. To construct the Hamiltonian for the max $d$-cut problem, the $d$-dimensional Clock matrices (specified in Equation \eqref{eq:clock}  \cite{anwar2012qutrit}) are used to appropriately replace the Pauli-$Z$ matrix in Equation \eqref{eq:IsingHamiltonian}, as was done for Max 3-Cut in Equation \eqref{eq:max3cutHamiltonian}. 

\begin{equation}
	V_{d}=\sum_{j=0}^{d-1}\ket{j}\bra{(j+1) \; mod \; d}
	\label{eq:shift}
\end{equation}
\begin{equation}
	U_{d}=\sum_{j=0}^{d-1}\omega^{j}\ket{j}\bra{j}
	\label{eq:clock}
\end{equation}
where  $\omega=e^{\nicefrac{2\pi i}{d}}$ is the $d^{th}$ root of unity \cite{anwar2012qutrit}.

The resultant Ising Hamiltonians for max $d$-cut is specified as:
\begin{equation}
	H=\sum_{ij}\frac{w_{ij}}{2}(U_{d}^{i}U_{d}^{j \dagger}+U_{d}^{i \dagger}U_{d}^{j})
	\label{eq:maxkcut}
\end{equation}

Further, it is possible to view the proposed method as a quantum mechanical extension of the vector Potts model \cite{Grimmett2006,wu1982potts}, where the spin-states of a $d$ level system are equally spaced on a unit circle, with the interaction energy of adjacent spins depending on the cosine of the relative angles between the states \cite{wu1982potts}. When the Ising Model contains \textit{external, longitudinal magnetic field} tems, Equantions \eqref{eq:IsingHamiltonian} and \eqref{eq:maxkcut} modify to Equations \eqref{eq:ising} and \eqref{eq:isingd}, respectively, where $h_i$ can be considered as an additional weight of the $i^{th}$ node of the graph under consideration. This is carried out by leveraging Equation \eqref{eq:1sigma}.

\begin{equation}
	\sigma_{z} = \frac{1}{2}(\sigma_{z} + \sigma_{z}^\dagger)
	\label{eq:1sigma}
\end{equation}

\begin{equation}
	H(\boldsymbol{\sigma})=\sum_{ij}w_{ij} \sigma_{z}^{i} \sigma_{z}^{j} + \sum_i h_i \sigma_{z}^i
	\label{eq:ising}
\end{equation}

\begin{equation}
	H=\sum_{ij}\frac{w_{ij}}{2}(U_{d}^{i}U_{d}^{j \dagger}+U_{d}^{i \dagger}U_{d}^{j}) + \sum_i \frac{h_i}{2}(U_d^i + U_d^{i \dagger})
	\label{eq:isingd}
\end{equation}

\begin{figure}
	\centering
	\includegraphics[scale=0.5]{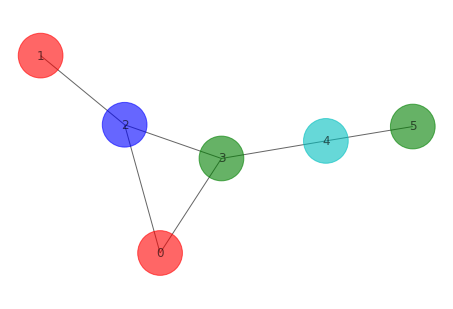}
	\caption{Results of max $4$-cut clustering on an unweighted graph where the four clusters are denoted by red, blue, green and cyan colours.}
	\label{fig:quaternary2}
\end{figure}

\begin{figure*}
	\centering
	\includegraphics[scale=0.63]{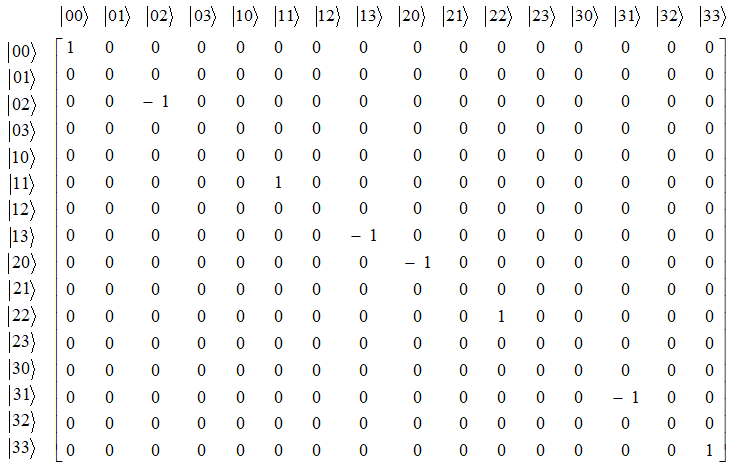}
	\caption{Node-node interaction for the quaternary clustering case}
	\label{fig:quaternaryHamiltonian}
\end{figure*}

A typical result of quaternary clustering for a simple graph is shown in Figure \ref{fig:quaternary2}. Again, extensive verification of the results of the formulation using different graphs and weights has been carried out.

For the $d=4$ case, the interaction Hamiltonian matrix for two adjacent nodes has been shown in Figure \ref{fig:quaternaryHamiltonian}. It should be noted that the $\ket{aa}$ elements are $1$, while the $\ket{ab}$ elements are $0$ or $-1$. This happens because there are two possible angles between the $4^{th}$ roots of unity, i.e., $\nicefrac{\pi}{2}$ and $\pi$. If the angle between the classes is $\nicefrac{\pi}{2}$, then the interaction term is 0, and it is -1 if the classes are $\pi$ radians apart. This means that having a larger angular difference between the classes is more favorable. The effect of clustering still remains the same, however. The system settles for a state that ensures the highest angular difference between the classes. This was not apparent for the $d=2$ or $d=3$ cases because there was only one possible angle between the classes. 

For a $d$-cluster problem, if the classes are numbered from $0$ to $(d-1)$, then the interaction energy term between nodes of classes $a$ and $b$ is given by:
\begin{equation}
	z_{a} \cdot z_{b}=\frac{1}{2}(\bar{z_{a}}z_{b}+z_{a}\bar{z_{b}})
	\label{dotproduct}
\end{equation}

where $z_{a}=e^{\nicefrac{2\pi ai}{d}}$ and $\bar{z_{a}}$ is the complex-conjugate of $z_{a}$.

\begin{figure}
	\centering
	\includegraphics[scale=0.65]{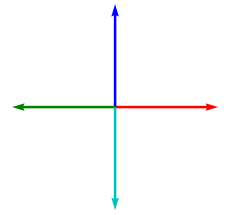}
	\caption{The cluster classes along the vectors: $0$ along red, $1$ along blue, $2$ along green, and $3$ along cyan.}
	\label{fig:quaternaryVector}
\end{figure}

Figure \ref{fig:quaternaryVector} shows the vector representation of the four classes along the fourth roots of unity. 

\begin{figure}
	\centering
	\includegraphics[scale=0.5]{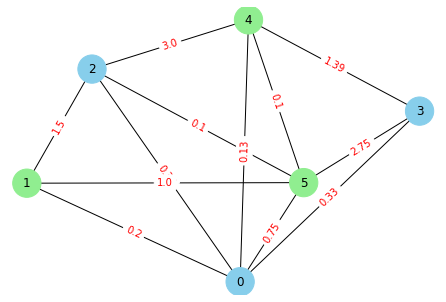}
	\caption{Results of max $4$-cut clustering where the nodes get placed into two clusters only, instead of four.}
	\label{fig:quaternary1}
\end{figure}

Additionally, for $d=4$, an interesting behaviour was observed.  For many graphs, the result was bipartite, i.e., the resultant partition had only two classes. The partitioning for such a graph has been shown in Figure \ref{fig:quaternary1}. Other graphs had solution states for $2$, $3$ and $4$ classes, all having the minimum energy eigenvalue.
This might have been due to the fact that adjacent classes have $0$ interaction energy. Further investigations are necessary for more than $4$ classes and to arrive at the nature of the behaviour for general $d$ classes.

\section{Proposition for Future Implentation of QUDO Algorithms} \label{implementation}
With the NISQ devices already being available, it is really essential to consider the implementation of the quantum algorithms on the hardware. Few pointers are put forth in this section. The binary max-cut problem can be solved using quantum annealing such that the system settles to the final Hamiltonian given by the Ising function. The D-Wave annealer Hamiltonian may be represented as \cite{dwave}:

\begin{equation}
	H_{ising}=-\frac{A(s)}{2}\left(\sum \sigma_{x}^{i}\right)+\frac{B(s)}{2}\left(\sum_{i} h_{i}\sigma_{z}^{i}+\sum_{i>j}J_{ij}\sigma_{z}^{i}\sigma_{z}^{j}\right)
	\label{eq:dwave}
\end{equation}

The cost function can easily be mapped to the D-Wave Ising Hamiltonian and the minimum energy states can be obtained by appropriate sampling.

Alternatively, the QAOA approach can also be utilised to find the solution of binary max-cut problem by evolving the appropriately prepared wavefunction using unitary operators towards the Hamiltonian obtained by casting the problem into Ising model (or the equivalent QUBO) \cite{hadfield2018quantum,zhou2018quantum}. Thus, QA and QAOA have formulation of the Ising model in common but they require different hardware. Ruslan Shaydulin, et al, have compared the two approaches for a different, but similar, problems \cite{shaydulin2018community}. In QAOA hybrid quantum and classical processing is utilised and the quantum computation is in terms of the gate-circuit model \cite{hadfield2018quantum,cervera2018exact}.

For max $d$-cut (or QUDO) problems, these approaches cannot be used directly. The existing
quantum annealers are inherently binary in nature, due to the presence of $2 \times 2$ Pauli matrices.
Ushijima-Mwesigwa, et al \cite{ushijima2017graph}, have proposed a concept of super nodes for Graph Partitioning into $d$
classes. The problem formulation is quite similar to that of max-cut, and the same
approach can be used for the latter. The drawback is that for a graph with $N$ vertices, $dN$ qubits
are required to model the problem onto an annealer and the matrix blows up by a factor of $d^{2}$.

If useful, future annealers can be designed in such a way that they utilise Equation \eqref{eq:maxkcut} as the
final Hamiltonian, with the addition of cross terms containing $U_{d}$ and $V_{d}$. Specifically, one can
think of annealers based on \emph{qutrits} \cite{Burlakov2003} to address max $3$-cut based on Equation \eqref{eq:max3cutHamiltonian}.

\subsection{Qudit-based Circuit of QAOA for Max $d$-Cut}
In this section, we elaborate on the steps to construct the gate-based quantum circuit for solving the problem of Max $d$-Cut using QAOA \cite{farhi2014quantum}, which utilizes a hybrid approach by leveraging the variational principle, for the $d$-ary case. In QAOA a $p$-layer ansatz $\ket{\psi(\boldsymbol{\gamma}, \boldsymbol{\beta})}$, with the cost and mixing Hamiltonians $H_c$ and $H_x$ are used, where:
\begin{equation}
	\ket{	\psi(\boldsymbol{\gamma}, \boldsymbol{\beta}) }= e^{-i \beta_p H_x}e^{-i \gamma_p H_c} \dots e^{-i \beta_1 H_x}e^{-i \gamma_1 H_c} \ket{+}^{\otimes n}
\end{equation}

The parameters $\boldsymbol\gamma$ and $\boldsymbol\beta$ are found using classical optimizers, subject to the minimization of $\braket{\psi(\boldsymbol{\gamma}, \boldsymbol{\beta})| H_c |\psi(\boldsymbol{\gamma}, \boldsymbol{\beta})}$. The modifications required to use QAOA on qudit-based systems to solve the Max $d$-Cut problem will be described in the following subsections.

\subsubsection{Preparation of the state $\ket{+}_d^{\otimes n}$}

It is widely known that $\ket{+}^{\otimes n}$ is used to refer to the zero-phase, equal-superposition state of the system, and is achievable by the application of the $H$ gate on all the qubits which are initially in the $\ket{0}$ state, i.e.:
\begin{equation}
	\ket{+}^{\otimes n} = H^{\otimes n} \ket{0}^{\otimes n}
\end{equation}

\begin{figure}[H]
	\centering
	\includegraphics[scale=0.65]{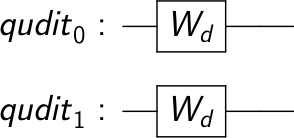}
	\caption{Circuit to place qudits in equal, zero-phase superposition from $\ket{0}_d$.}
	\label{fig:qudit_superposition}
\end{figure}

To obtain a similar superposition for qudits, we propose the use of the generalized Walsh-Hadamard matrix $W_d$ \cite{sylvester}, in a $d$-dimensional Hilbert space, where $\omega = e^{\sfrac{2\pi i}{d}}$. To distinguish qudits from the traditionally used qubits, we represent the qudit states as $\ket{x}_d$. Figure \ref{fig:qudit_superposition} shows the circuit to obtain the state $\ket{+}_d$ from $\ket{0}_d$.

\begin{equation}
	W_d = \frac{1}{\sqrt{d}} 
	\begin{bmatrix}
		1 & 1 & 1 &\dots& 1 \\
		1 & \omega^{d-1} & \omega^{2(d-1)} &\dots& \omega^{(d-1)^2}\\
		1 & \omega^{d-2} & \omega^{2(d-2)} &\dots& \omega^{(d-1)(d-2)}\\
		\vdots & \vdots & \vdots &\ddots& \vdots \\
		1 & \omega & \omega^2 & \dots & \omega^{d-1} \\
	\end{bmatrix}
\end{equation}

\begin{equation}
	\ket{+}_d^{\otimes n} = W_d^{\otimes n} \ket{0}_d^{\otimes n}
\end{equation}

\subsubsection{Construction of $e^{-i \gamma_m H_c}$}

For the given max $d$-cut unitary to be implemented using qudit-circuits, it is imperative to define the following new unitary gates:

A two-qudit controlled-$V_d$ gate that applies the $V_d^k$ gate on the target-qudit when the control-qudit is in the state $|k\rangle_d$. This gate is analogous to the traditional CX (Controlled-NOT) gate in the qubit case. The matrix-form of the gate, with the $i^{th}$ qudit as control, and $j^{th}$ qudit as the target, is given by $CV_d^{(ij)} = diag(I,V_d,V_d^2,\dots,V_d^{d-1})$, where $V_d = \sum_{l=0}^{d-1} |l \rangle \langle (l+1) \; mod \; d|$ and $V_d^k = \sum_{l=0}^{d-1} |l \rangle \langle (l+k) \; mod \; d|$, which extends from Equation \eqref{eq:shift}.

The other gate $\mathcal{U}_d$ takes $d$ parameters and is the qudit counterpart of the $R_z$ gate; $R_z(\theta)$ is a single-qubit rotation through angle $\theta$ (radians) around the $z$-axis. If the parameters are given by $\Phi = \{\phi_0, \phi_1, \dots, \phi_{d-1} \}$, then $\mathcal{U}_d = diag(e^{i\phi_0}, e^{i\phi_1}, \dots, e^{i\phi_{d-1}})$. It may immediately be observed that $\mathcal{U}_d = U_d$, when $\Phi = \{0, \frac{2\pi}{d}, \frac{3\pi}{d}, \dots, \frac{2(d-1)\pi}{d} \}$.

\begin{figure*}
	\centering
	\includegraphics[scale=0.6]{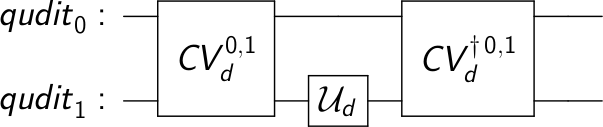}
	\caption{Circuit to apply the $e^{-i \gamma_m H_c}$ operator  on qudits $0$ and $1$ for an edge between the corresponding nodes.}
	\label{fig:qudit_oracle}
\end{figure*}

In the $d$-ary case for max-cut, discussed in Section \ref{maxkcut}, each node in the graph is represented by a qudit in the quantum circuit. In the venture to solve max $d$-cut using QAOA, the Ising Hamiltonian derived in Equation \eqref{maxkcut} functions as $H_c$, from which the evolution operator  $e^{i\gamma_m H_c}$ is constructed as shown in Equation \eqref{eq:qaoa_oracle}. The corresponding circuit for a single edge between the nodes $0$ and $1$ of a graph has also been portrayed in Figure \ref{fig:qudit_oracle} with parameter $\gamma_m$ for the $m^{th}$ layer of QAOA. The same argument can be extended to form the cost evolution operator for every edge in the graph and each layer.
\begin{gather}
	e^{-i \gamma_m H_c} = 
	e^{-i \gamma_m \sum_{ij} \frac{w_{ij}}{2}(U_dU_d^\dagger+U_d^\dagger U_d)} = \nonumber \\
	\prod_{ij} CV_d^{(ij)} \cdot I \otimes \mathcal{U}_{d\Phi} \cdot CV_d^{\dagger (ij)} 
	\label{eq:qaoa_oracle} 
\end{gather}
where \\
$\Phi = \{-\gamma_m w_{ij}, -\gamma_m w_{ij} cos(\sfrac{2\pi}{d}), \dots, -\gamma_m w_{ij} cos(\sfrac{2(d-1)\pi}{d})\} $.

\subsubsection{Construction of Mixing Operator, $e^{-i \beta_m H_x}$}

QAOA, being a trotterized, adiabatic annealing procedure, relies on the use of a mixing or driver Hamiltonian $H_x$. Traditionally, the single-qubit version of Grover's Diffusion Operator has been used as the mixing operator. This is evident from the fact that $e^{i\sfrac{\beta X}{2}}= R_x(-\beta) = H((1-e^{i\beta})|0\rangle \langle0| - I)H$, upto a global phase factor \cite{akshay};  where $R_x (\theta)$ signifies a rotation of the state about the $x$-axis by an angle of $\theta$ radians. With various extensions of the QAOA algorithm, better mixing Hamiltonians -- which are more suited to the problem at hand -- have been proposed \cite{quantum_alternate}. However, in this implementation, we will consider the single-qudit version of diffusion operator:
\begin{equation}
	e^{-i\beta_m H_x} = \prod_j W_d \cdot \mathcal{U}_{d(-2\beta_m, 0, \dots, 0)} \cdot W_d^\dagger
\end{equation}

\begin{figure}[H]
	\centering
	\includegraphics[scale=0.45]{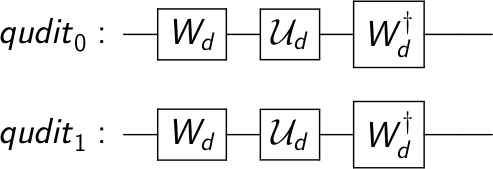}
	\caption{Circuit to implement the $e^{-i \beta_m H_x}$ operator}
	\label{fig:qudit_mixing}
\end{figure}

The circuit in Figure \ref{fig:qudit_mixing} shows the procedure to apply the mixing operator in the $m^{th}$ layer to two of the qudits with parameter $\beta_m$. Again, the same circuit with corresponding parameters can be used for ach of the $p$ layers in QAOA. The absence of qudit-based quantum processors and simulators has deterred us from practically trying out the circuits. However, as mentioned in earlier sections, the numerical evaluation of the results based on the suggested formulation did bring out the usefulness of the approach.

This concludes the discussion related to both annealer and gate-model based implementation suggestions for QUDO formalism.

\section{Conclusions}\label{conclusions}

The paper, by appropriately combining mathematical aspects and supporting numerical results proposed quantum-assisted graph clustering for three or more clusters. Even though the presentation is biased towards algorithmic aspects, pointers are provided for possible architectures for implementation. In the process, qudit based circuit for max $d$-cut pivoted on Quantum Approximate Optimization Algorithm is worked out. While firm footing has been established for max $3$-cut case, for more than $3$ clusters one can carry out additional research to refine the clustering results.

\section*{Acknowledgment}

The authors sincerely thank Mr. Mahesh Rangarajan, Dr. Arpan Pal and Dr. Balamuralidhar P of TCS R\&I for their support and encouragement.

\begin{appendices}

\section{Additional Results Related to Max-Cut}
This Appendix demonstrates how quantum description/formulation can
sometimes lead to a different perspective and way of solving
problems by providing the results for subgraph identification in
graphs. Let $G(V,E)$ be a graph with $V$ as the set of vertices, having cardinality $N$, and $E$ as the set of edges. Max-cut partitioning allows us to cluster the set of vertices into two subsets of $V$, subject to minimising the cost function. It may happen that there are multiple solutions corresponding to the minimum cost. Then these degenerate solutions can be used to further partition the graph’s vertices into $M$ sets, $\bar{V_{1}}$, $\bar{V_{2}}$, and so on, where $V_{i} \in V$, and $M<N$. The clustering of the vertices in $\bar{V_{i}}$ does not influence the clustering of the vertices in any other set $\bar{V_{j}}$. The edges between the vertices of the set $\bar{V_{i}}$ forms the set $\bar{E_{i}}$, and the edges connecting vertices of $\bar{V_{i}}$ to that of set $\bar{V_{1}}$ are discarded, where $i \neq j$. Thus $\bar{V_{i}}$ and $\bar{E_{i}}$ together form a graph $\bar{G_{i}}$ which is a subgraph of the original graph $G$. The set of these subgraphs can be called independent subgraphs of $G$ as the max-cut clustering of $\bar{G_{i}}$ has no effect on the clustering of $\bar{G_{i}}$, given $i \neq j$.

The dual of a binary number is given by converting the $0$s to $1$s and $1$s to $0$s. Dual binary numbers, thus, represent equivalent partitions of the graph into two clusters, and will have the same energy or cost value. But if there exist solutions that are not duals of each other, and yet have the same clustering cost, then the graph can be said to have independent subgraphs within it. It can easily be seen that if the number of such independent subgraphs is $L$, then there will be $2^{L}$ solutions with the minimum eigenvalue.

For example, if the minimum energy computational states are:

$0011010          \;            -\{1\}$

$0011110            \;          -\{2\}$

$1100001              \;        -\{3\}$

$1100101                \;      -\{4\}$

\{1\}and \{4\} are duals of each other, as are \{2\} and \{3\}. They represent the same partitions of the graph. Though \{1\} and \{2\} have the same energy, they represent different partitions. Thus, it does not change the cost function whether node 4 is classified as into class $0$ or $1$. Similarly with \{3\} and \{4\}. 

The bitwise XOR of \{1\} and \{2\} gives $1$ on the fifth place from the left. As the bits in the solution refer to the clusters that the nodes of the graph are placed in, solutions \{1\} and \{2\} place all the nodes in similar clusters, except for the fourth node. Thus, the node 4 can be said to be independent of the rest of the graph, as their respective clustering do not affect each other, as well as the overall cost function. The graph then contains $2$ idependent subgraphs: $G_{1}$ and $G_{2}$, having vertex sets $V_{1}=\{0,1,2,3,5,6\}$ and $V_{2}=\{4\}$, respectively. 

\begin{figure}[h]
	\centering
	\begin{subfigure}[b]{0.4\textwidth}
		\centering
		\includegraphics[width=\textwidth]{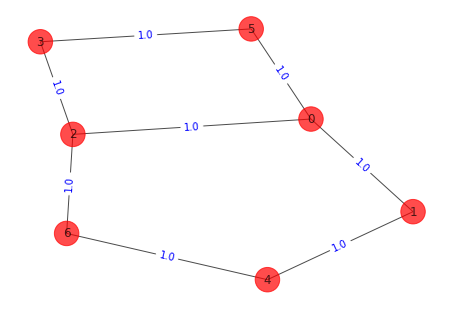}
		\caption{Input graph}
		\label{fig:independentInput}
	\end{subfigure}
	\hfill
	\begin{subfigure}[b]{0.4\textwidth}
		\centering
		\includegraphics[width=\textwidth]{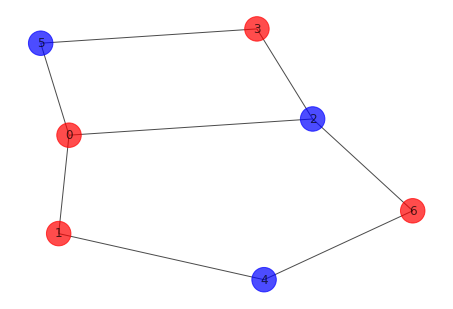}
		\caption{Output after max-cut partitioning}
		\label{fig:independentOutput}
	\end{subfigure}
	\hfill
	\caption{Example graph to show independent subcomponents}
	\label{fig:independent}
\end{figure}

Thus, the non-transverse Ising Hamiltonian gives the independent and isolated subgraphs of a graph. A different cost function will provide different subgraphs corresponding to some other property represented by that cost function. This can be visualised by the graph shown in Figure \ref{fig:independent}. On max-cut partitioning the graph in \ref{fig:independentInput}, the graph \ref{fig:independentOutput} is obtained, with the nodes clustered into either the red or blue cluster.

The solutions with the minimum energy values are given by:

$0010110	\;-\{1\}$

$0110010	\;-\{2\}$

$0110011	\;-\{3\}$

$0110110	\;-\{4\}$

$1001001	\;-\{5\}$

$1001100	\;-\{6\}$

$1001101	\;-\{7\}$

$1101001	\;-\{8\}$

The solutions \{5\}, \{6\}, \{7\} and \{8\} are the duals of \{4\}, \{3\}, \{2\} and \{1\}, respectively, and thus, do not incorporate any extra information about the partitioning of the system. The bitwise XOR of \{1\} and \{2\} gives nodes 1 and 4; \{1\} and \{3\} gives 1, 4 and 6; \{1\} and \{4\} gives 1; \{2\} and \{3\} gives 6; \{2\} and \{4\} gives 4; and \{3\} and \{4\} gives nodes 4 and 6. The super set of all the resulting vertices $\{1,4,6\}$ consitutes a subgraph $G1$ whose partitioning does not influence that of the remaining graph. It can further be inferred from the results that $G1$ can further be partitioned into subgraphs $\{1\}$ and $\{4,6\}$, or $\{1,4\}$ and $\{6\}$.

\end{appendices}


\begin{thebibliography}{99}
	
	\bibitem{2018arXiv181203050T}
	L.~{Tse}, P.~{Mountney}, P.~{Klein}, and S.~{Severini}, ``{Graph cut
		segmentation methods revisited with a quantum algorithm},'' {\em arXiv
		e-prints}, p.~arXiv:1812.03050, Dec 2018.
	
	\bibitem{kim2019leveraging}
	M.~Kim, D.~Venturelli, and K.~Jamieson, ``Leveraging quantum annealing for
	large mimo processing in centralized radio access networks,'' in {\em
		Proceedings of the ACM Special Interest Group on Data Communication},
	pp.~241--255, ACM, 2019.
	
	\bibitem{preskill2018quantum}
	J.~Preskill, ``Quantum computing in the nisq era and beyond,'' {\em Quantum},
	vol.~2, p.~79, 2018.
	
	\bibitem{1997quant.ph..1001B}
	C.~H. {Bennett}, E.~{Bernstein}, G.~{Brassard}, and U.~{Vazirani}, ``{Strengths
		and Weaknesses of Quantum Computing},'' {\em arXiv e-prints},
	pp.~quant--ph/9701001, Jan 1997.
	
	\bibitem{zhu2019training}
	D.~Zhu, N.~M. Linke, M.~Benedetti, K.~A. Landsman, N.~H. Nguyen, C.~H.
	Alderete, A.~Perdomo-Ortiz, N.~Korda, A.~Garfoot, C.~Brecque, {\em et~al.},
	``Training of quantum circuits on a hybrid quantum computer,'' {\em Science
		advances}, vol.~5, no.~10, p.~eaaw9918, 2019.
	
	\bibitem{otterbach2017unsupervised}
	J.~Otterbach, R.~Manenti, N.~Alidoust, A.~Bestwick, M.~Block, B.~Bloom,
	S.~Caldwell, N.~Didier, E.~S. Fried, S.~Hong, {\em et~al.}, ``Unsupervised
	machine learning on a hybrid quantum computer,'' {\em arXiv preprint
		arXiv:1712.05771}, 2017.
	
	\bibitem{lewis2017quadratic}
	M.~Lewis and F.~Glover, ``Quadratic unconstrained binary optimization problem
	preprocessing: Theory and empirical analysis,'' {\em Networks}, vol.~70,
	no.~2, pp.~79--97, 2017.
	
	\bibitem{farhi2014quantum}
	E.~Farhi, J.~Goldstone, and S.~Gutmann, ``A quantum approximate optimization
	algorithm,'' 2014.
	
	\bibitem{1525}
	R.~(https://physics.stackexchange.com/users/369/raskolnikov), ``Ising model for
	dummies.'' Physics Stack Exchange.
	\newblock URL:https://physics.stackexchange.com/q/1525 (version: 2010-12-01).
	
	\bibitem{kamenetsky2010ising}
	D.~Kamenetsky {\em et~al.}, ``Ising graphical model,'' 2010.
	
	\bibitem{atan}
	A.~Tan, ``Quantum ising models,'' 2018.
	
	\bibitem{ushijima2017graph}
	H.~Ushijima-Mwesigwa, C.~F. Negre, and S.~M. Mniszewski, ``Graph partitioning
	using quantum annealing on the d-wave system,'' in {\em Proceedings of the
		Second International Workshop on Post Moores Era Supercomputing}, pp.~22--29,
	ACM, 2017.
	
	\bibitem{gowda}
	T.~Gowda, ``Introduction to quantum optimization using {D-W}ave {2X},'' 2018.
	
	\bibitem{Alan}
	A.~Frieze and M.~Jerrum, ``Improved approximation algorithms for max k-cut and
	max bisection,'' in {\em Integer Programming and Combinatorial Optimization}
	(E.~Balas and J.~Clausen, eds.), (Berlin, Heidelberg), pp.~1--13, Springer
	Berlin Heidelberg, 1995.
	
	\bibitem{prakash2018normal}
	S.~Prakash, A.~Jain, B.~Kapur, and S.~Seth, ``Normal form for single-qutrit
	clifford+ t operators and synthesis of single-qutrit gates,'' {\em Physical
		Review A}, vol.~98, no.~3, p.~032304, 2018.
	
	\bibitem{Frydryszak_2017}
	A.~Frydryszak, L.~Jakóbczyk, and P.~Ługiewicz, ``Determining quantum
	correlations in bipartite systems - from qubit to qutrit and beyond,'' {\em
		Journal of Physics: Conference Series}, vol.~804, p.~012016, Jan 2017.
	
	\bibitem{anwar2012qutrit}
	H.~Anwar, E.~T. Campbell, and D.~E. Browne, ``Qutrit magic state
	distillation,'' {\em New Journal of Physics}, vol.~14, no.~6, p.~063006,
	2012.
	
	\bibitem{Grimmett2006}
	G.~Grimmett, {\em The Random-Cluster Model}, ch.~10, pp.~320--340.
	\newblock Springer Berlin Heidelberg, 2006.
	
	\bibitem{wu1982potts}
	F.-Y. Wu, ``The potts model,'' {\em Reviews of modern physics}, vol.~54, no.~1,
	p.~235, 1982.
	
	\bibitem{dwave}
	``{D-W}ave system documentation.''
	
	\bibitem{hadfield2018quantum}
	S.~Hadfield, ``Quantum algorithms for scientific computing and approximate
	optimization,'' {\em arXiv preprint arXiv:1805.03265}, 2018.
	
	\bibitem{zhou2018quantum}
	L.~Zhou, S.-T. Wang, S.~Choi, H.~Pichler, and M.~D. Lukin, ``Quantum
	approximate optimization algorithm: performance, mechanism, and
	implementation on near-term devices,'' {\em arXiv preprint arXiv:1812.01041},
	2018.
	
	\bibitem{shaydulin2018community}
	R.~Shaydulin, H.~Ushijima-Mwesigwa, I.~Safro, S.~Mniszewski, and Y.~Alexeev,
	``Community detection across emerging quantum architectures,'' {\em arXiv
		preprint arXiv:1810.07765}, 2018.
	
	\bibitem{cervera2018exact}
	A.~Cervera-Lierta, ``Exact ising model simulation on a quantum computer,'' {\em
		Quantum}, vol.~2, p.~114, 2018.
	
	\bibitem{Burlakov2003}
	A.~V. Burlakov, L.~A. Krivitskii, S.~P. Kulik, G.~A. Maslennikov, and M.~V.
	Chekhova, ``Measurement of qutrits,'' {\em Optics and Spectroscopy}, vol.~94,
	p.~684–690, May 2003.
	
	\bibitem{sylvester}
	J.J. Sylvester (1867) LX. ``Thoughts on inverse orthogonal matrices, simultaneous sign successions, and tessellated pavements in two or more colours, with applications to Newton's rule, ornamental tile-work, and the theory of numbers,'' 
	{\em The London, Edinburgh, and Dublin Philosophical Magazine and Journal of Science}, 34:232, 461–475.
	
	\bibitem{akshay}
	V.~Akshay, H.~Philathong, M.E.S.~Morales, and J.~Biamonte, ``Reachability deficits in quantum approximate optimization,''
	{\em Physical Review Letters}, vol.~124, Mar 2020.
	
	\bibitem{quantum_alternate}
	S.~Hadfield, Z.~Wang, Br.~O'Gorman, E.G.~Rieffel, D.~Venturelli, and R.~Biswas, ``From the quantum approximate optimization algorithm to a quantum alternating operator ansatz,''
	{\em Algorithms}, vol.~12, Feb 2019.
	
	
	
\end{thebibliography}
\end{document}